\documentclass[10pt]{amsart}
\usepackage{geometry}       
\usepackage{subfig}
\usepackage[english]{babel}
\usepackage{authblk}
\usepackage{todonotes}
\usepackage{amsaddr}
\usepackage[latin1]{inputenc}
\usepackage{amsmath,amssymb,latexsym,amsthm}
\numberwithin{equation}{section}
\usepackage{float}
\usepackage{enumerate}
\usepackage{fancyhdr}
\usepackage{url}

\makeatletter
\g@addto@macro{\endabstract}{\@setabstract}
\newcommand{\authorfootnotes}{\renewcommand\thefootnote{\@fnsymbol\c@footnote}}%
\makeatother

\title[A validation of data sciences use in mines]{A validation of the use of data sciences for the study of slope stability in open pit mines}
\begin{document}

\maketitle

\begin{center}
  \normalsize
    \bigskip\footnote{Email: jortega@dim.uchile.cl, mrapiman@gesecologygroup.com, lrojo@gesecologygroup.com. 
This work was partially supported by gesecology Group Ltda. and CORFO 16ITE2-60290 Geoalert 2.0.}
  \authorfootnotes
  J. H. Ortega\textsuperscript{2}, M. Rapiman\textsuperscript{1}
  L. Rojo\textsuperscript{1},  J. P.Rivacoba\textsuperscript{1}
   \par \bigskip
  \textsuperscript{1}Gesecology Group Ltda., Los Andes, Chile\par
  \textsuperscript{2}Departamento de Ingenier\'ia Matem\'atica, Universidad de Chile and 
Center for Mathematical Modeling (CMM), Universidad de Chile, Santiago, Chile\par 
 
\end{center}

\begin{abstract}
In this work, we present an exploratory study of stability of an open pit mine in the north of Chile with the use of data mining. It is important to note that the study of slope stability is a subject of great interest to mining companies, this is due to the importance in the safety of workers and the protection of infrastructures, whether private or public, in those places susceptible to this kind of phenomena, as well as, for road slopes and close to communities or infrastructures, among others. It is also important to highlight that these phenomena can compromise important economic resources and can even cause human losses. In our case, this study seeks to increase the knowledge of these phenomena and thus, try to predict their occurrence, by means of risk indicators, potentially allowing the mining company supervision to consider predictive measures. It should be considered that there is no online test that ensures timely prediction.  In previous studies conducted in other mines, it has been corroborated that the phenomena and factors associated with the movement of slopes and landslides are extremely complex and highly nonlinear, which is why the methods associated with the called {\it data mining}, were found to be ideal for discovering new information in the data, which is recorded periodically in the continuous monitoring that mining companies have of their deposits, which allowed to find important correlations for the search of predictors of these phenomena. Some results, with data coming from different sources, are presented at the end of this work.  We note that according to the information provided by the mining company, the results were favorable in the indicators of up to six months, giving as risk areas the correct sectors and predicting the April 2017 event. 
\end{abstract}

\setcounter{page}{2} 

\section{Introduction}
\
\par

The movement of the slopes is a common phenomenon in open pit mines. The slopes of these mines are designed with a safety factor to control the risk of injury and equipment damage due to slope failure and rock fall. In what follows we will present a brief bibliographical review about the precursors of slope instabilities. Dubrovskiy and Sergeev \cite{Dubrovskiy2006}, point out that there is the possibility of finding phenomena that predict the unstable state of a slope and for which a collapse of the slope can be predicted in time and scale. Slopes never fail spontaneously. Prior to failure, the slope provides indications in the form of measurable movement and\/ or the appearance of tension cracks. In contrast to this, the movements that present the slopes, are the result of the long-term movement of these, which have dragged for hundreds of years resulting in the cumulative movement of tens of meters. Precursor movements and early indicators can warn landslide disasters \cite{Rai2014}. As stated by this author, there are different ways of deformation and faults that can happen in a pronounced slope, but common to all of them, when a slope is excavated or exposed, there is an initial period of response as a result of stress relaxation \cite{Zavodni2000}, being most common in mines that have a rapid excavation rate. Reeves et al. \cite{Reeves2000}, studied the detection of precursor movements of slopes, using a system with real radar interferometry, the SSR system (Slope Stability Radar). This system was later commercialized, through the company Ground Probe in 2001. The hypothesis that was handled is that a system would be able to detect these very small precursor movements that occur prior to the collapse of the slope, which would allow estimating the stability of the slope under study \cite{Reeves2000}. Note that this is valid for fragile clumps, however other clumps allow to accumulate deformation in volumes of different scale without generating a violent failure, but a permanent accommodation.  It should be noted that the rain seems to increase the movements of the slope, with the water pressure acting to destabilize it even more \cite{Reeves2000}. Faillettaz and Or \cite{Faillettaz2013}, point out that, being able to identify that one is facing an imminent failure by gravity in a natural environment (for example a slope), is a complex and daunting task, especially because sudden ruptures are processes highly nonlinear, sensitive to unknown heterogeneities, inherent in natural media.

	These authors postulate that these mechanical failures release elastic energy measurable as microsismos or acoustic emissions. Thus, by monitoring these activities they should provide information about the mechanical conditions of the faults, however, the occurrence of the catastrophic rupture in heterogeneous materials is not an instantaneous event, but is typically preceded by smaller internal faults before the breaking off. It should be noted that microseismics is a powerful tool, but it has not been massively used in mining processes in bad to regular massifs. There are no documents about it or enough experience in geomechanical knowledge. The prediction in bursts of rock is still something that is not fully understood or predicted, but which is a field of great interest and study in underground mining.   The results show that the failure mode can vary dramatically with the redistribution of the charge from a diffuse damage to the growth of a single crack. These changes affect the statistical properties of the ruptures of the preceding micro-cracks. While the behavior of diffuse damage, it exhibits clear precursory signs (such as the increase in seismic activity before the definitive rupture). The individual behavior of the growth of a crack leads to a sudden failure, without being preceded by obvious precursors.  There are also slope geomorphological precursors that reflect areas of risk of earth movements, as occurred in the Chi-chi earthquake, a small town in Central Taiwan \cite{Wang2003}. This earthquake (7.9 Richter) triggered a huge landslide of 102 hectares, the {\it Chiu-fen-erh-shan}, of 50 million cubic meters, and which were deposited on 92.5 hectares of arable land. Analysis of aerial images one year before the event, showed anomalous characteristics, such as, a linear depression, zones of deep depression, a steep passage, and a low drainage density. All this added to the geological characteristics of the zone, first, the zone presents / displays a clay bed that covers all the zone, and second, the slope of the slope goes growing downwards, and on the other hand there was a bed of sandstone that supported this structure, and that was removed during the earthquake.

	Hartwig et al. \cite{Hartwig2013} used PSI (Persistent Scatterer Interferometry) techniques to monitor an iron mine in Carajás, Brazil, using 18 scenes of TerraSAR-X acquired during the dry season. Although, most of the study area was stable, high rates of deformation were detected in dumps (312 mm / year). Although geotechnical designs can be improved to increase safety factors and bank designs can be improved to minimize the hazards of falling rocks. However, even on slopes with conservative designs, they may experience unexpected failures, due to the presence of unknown geological structures, abnormal atmospheric states, or seismic movements \cite{Osasan2012}.  For slope monitoring, a series of systems capable of identifying and monitoring the precursors of greater slope collapses or rock mass movement have been developed. SqueeSAR was used to confirm the movement of an area of the Italian Alps, before and after some events were triggered, but with constant and relatively low movement rates \cite{Notti2013}. In addition to detecting and following slope instabilities, due to its ability to detect small movements. In \cite{Kayesa2006} the authors   used Geomos (Geodetic Monitoring System), a Lasica system from Leica Geosystems to monitor a gold mine (Letlhakane mine) in Botswana, Africa, to predict the collapse of the slope of the mine. This collapse was preceded by precursor movements of tension crack formation, The extension of these cracks, as well as the increase of the movement itself of the slope, which was registered with the Geomos system. The system predicted the impending failure, although not the exact moment of the event.

	In \cite{Szwedzicki2001} the authors points out that the analysis of the collapse events in open-pit and underground mines throws evidence of the existence of precursors of rock mass failure. Their results showed that there is a sequence that leads to the collapse of the rock mass. The authors point out that the movements of the slope were observed in the periphery of each impending collapse during prolonged periods, in some cases, for years. They shows various geotechnical precursors of several mine collapses in the world. Of these cases, it analyzes the collapse of the open gold mine wall (Telfer mine), located in the western part of Australia. They also note that the study of documented cases of large-scale mass movements indicates that the collapse does not happen without warning. Structural failures are manifested by the presence of geotechnical precursors. The precursors warn the development of large deformations of the terrain or high tensions. There is a series of precursors that are registered in the field, in time, intensity and location. According to the cases studied, Szwedzicki (2001) indicates the sequence of the precursors that have been registered: 
\begin{itemize}
\item Years before: earth movements occur in the periphery of the collapse zone. These movements refer, in general, to surface cracking, cracks that open, or displacement of the rock mass, either vertical or horizontal.  

\item Months before: the precursors are more localized and can be broadly described as deteriorating ground conditions. Notable precursors include subsidence of the surface, small occurrences of faults in the field.  

\item Weeks earlier: in hard rock mines, the precursors are seen close to the neighborhood of the center of the collapse zone. The main precursors are landslides and rock falls.  Hours before: the most common observations are rock noises and large landslides. These precursors appear in the center of the zone to collapse.
\end{itemize}

\section{Data Analysis and results }

This project was developed in an open pit mine located in the north region of Chile. The study region corresponds to the a bank  where there was an instability event during 2016. The area is shown in the image below.

\begin{figure}
\begin{center}
\includegraphics[width=5in]{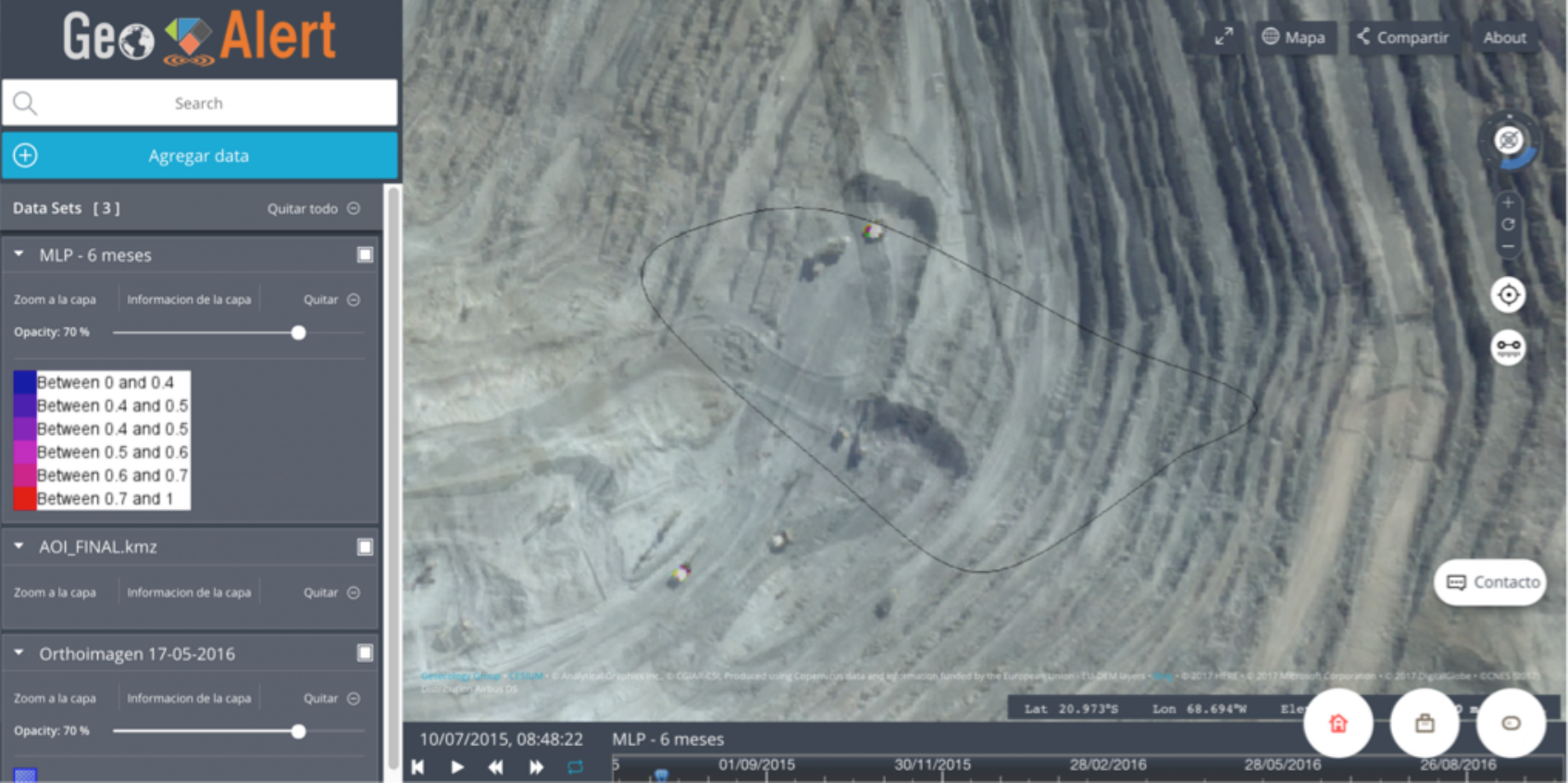}
\caption{Study zone of the mine }
\label{Fig1}
\end{center}
\end{figure}

\smallskip\

For this study, the following sources of information were available:  
\begin{itemize}
\item Prisms 
\item	Piezometers 
\item	Block Model 
\item	Digital elevation models
\end{itemize}

\begin{figure}
\begin{center}
\includegraphics[width=5in]{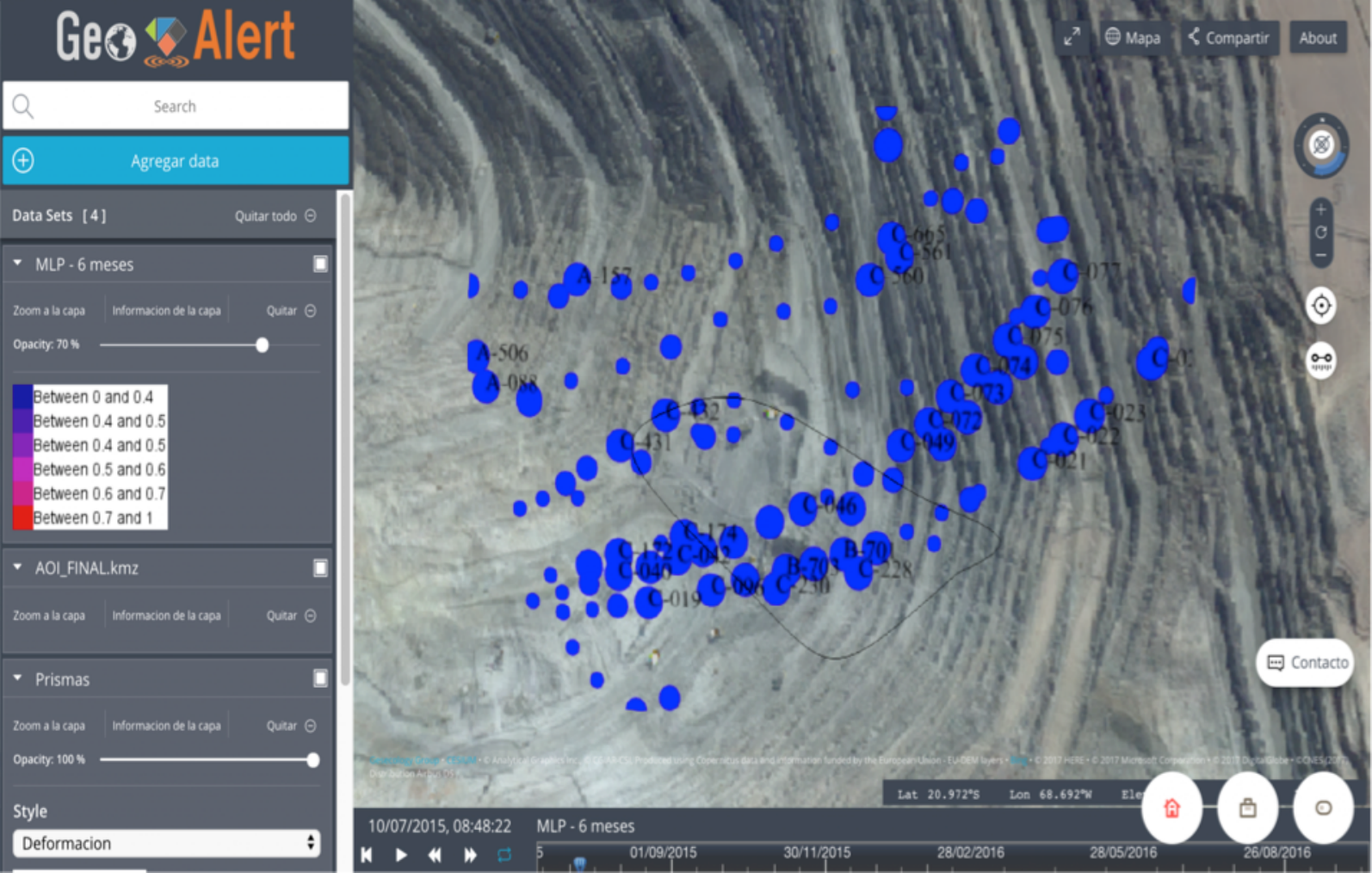}
\caption{Distribution of prims in the study region (GEOALERT)}
\label{Fig2}
\end{center}
\end{figure}

In Figure \ref{Fig1} we can see the region of study. The position of the delivered prism data is shown in Figure \ref{Fig2}.

\smallskip\

The following prisms information were available: 
\begin{itemize}
\item	116 prisms
\item	9980 displacement measurements in total, evenly distributed for the period sent
\item	Data for 2014 - 2015 - 2016,
\end{itemize}

\smallskip\

\begin{figure}
\begin{center}
\includegraphics[width=5in]{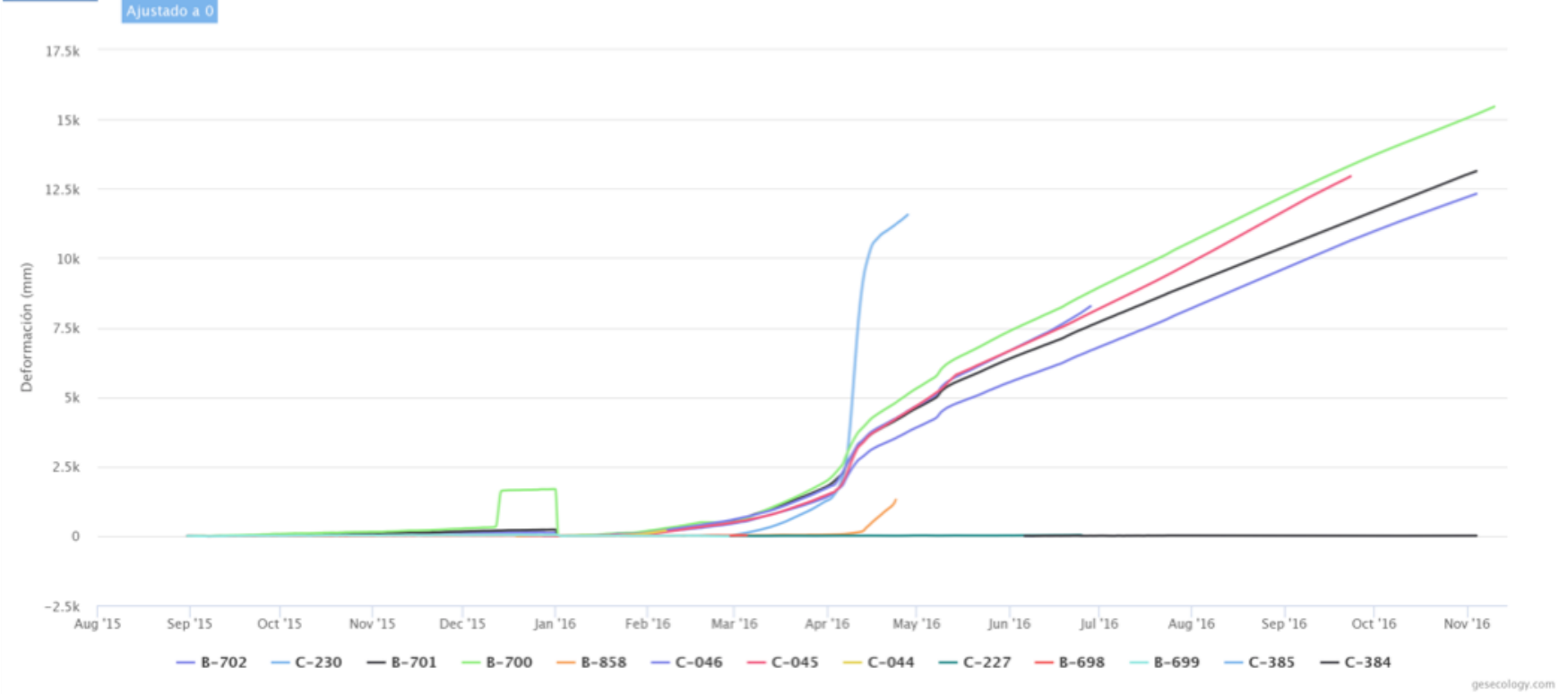}
\caption{Deformation data obtained from prisms network}
\label{Fig3}
\end{center}
\end{figure}

In Figure \ref{Fig3}, time series of deformation of the prisms that are within the study area are shown.

\smallskip

\begin{figure}
\begin{center}
\includegraphics[width=5in]{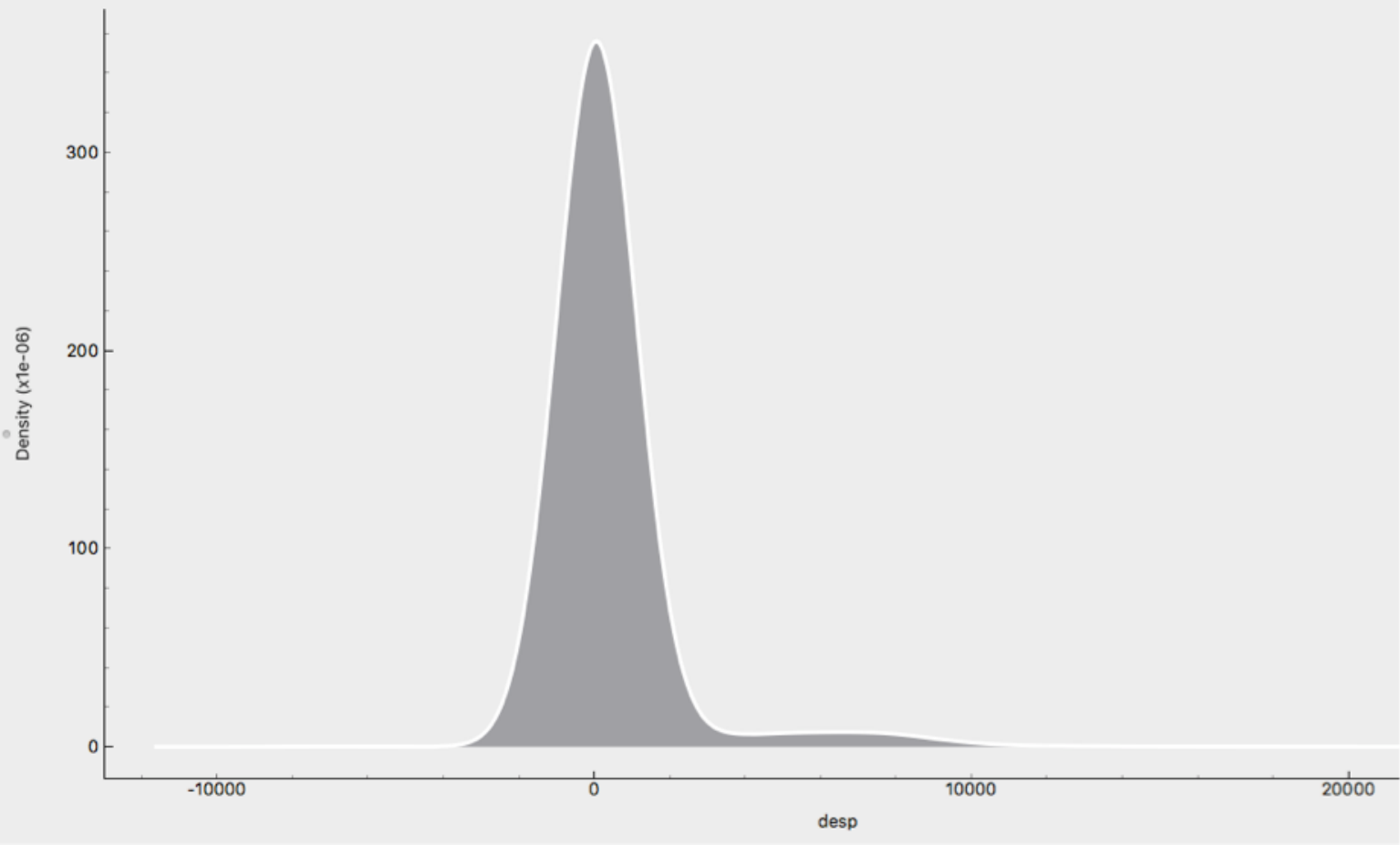}
\caption{Deformation data obtained from prisms network}
\label{Fig4}
\end{center}
\end{figure}

The distribution of the measurements is showed in Figure \ref{Fig4}.

For the data of piezometers, the following information was available: 
\begin{itemize}
\item	45 wells
\item On average, 20,000 pressure measurements
\item In some cases, data from 2010 to 2016
\end{itemize}
Finally, the block model consisted of several fields that were included in the model.

\section{Risk assessment model results}
\
\par

\begin{figure}
\begin{center}
\includegraphics[width=5in]{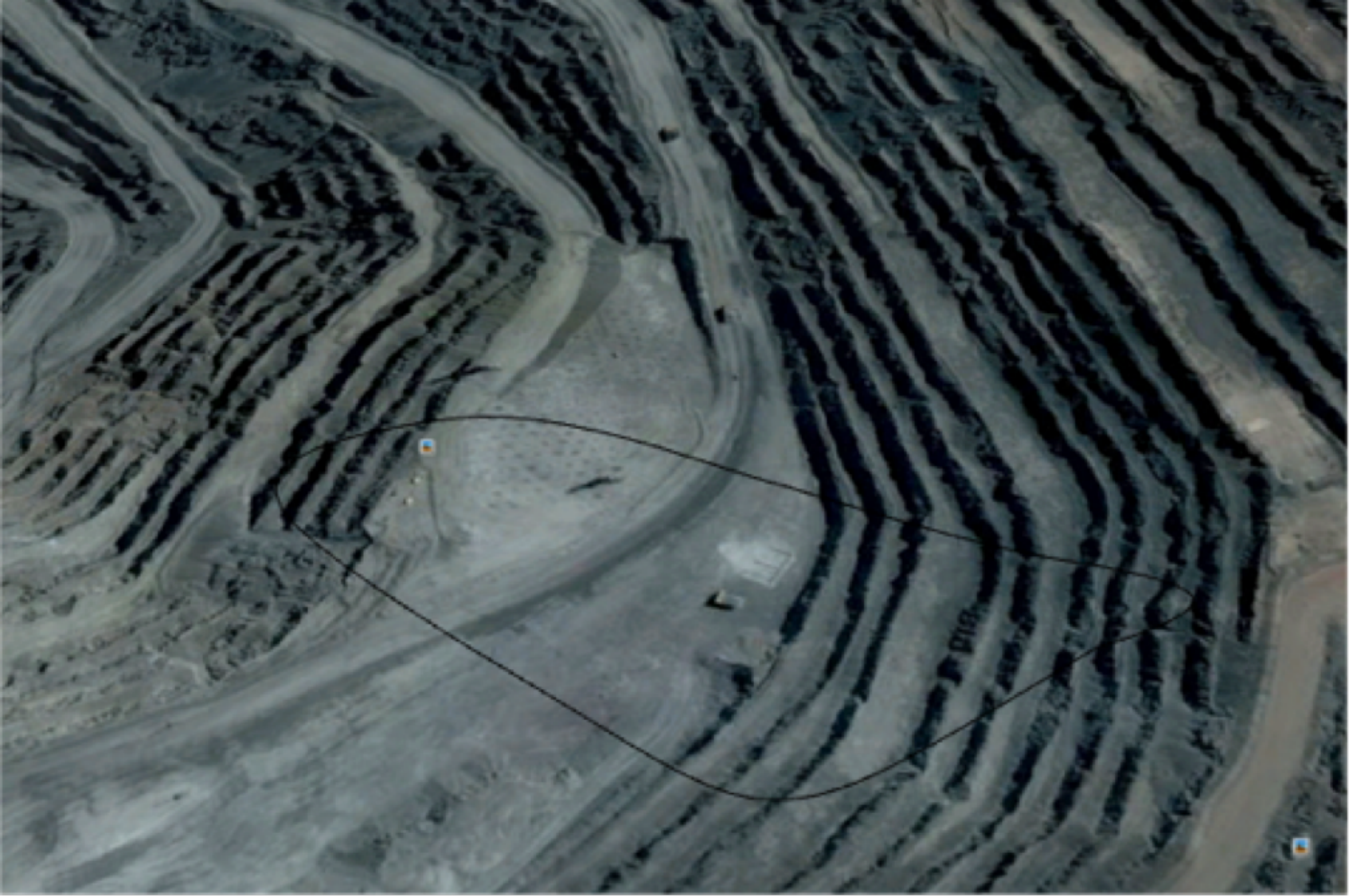}
\caption{Study zone of the mine}
\label{Fig5}
\end{center}
\end{figure}

In what follow, we will present the results delivered by the predictive model. Figure \ref{Fig5} shows the polygon considered for the study, we note that in that region we have information about a know instability, this zone is located in the SE sector of the mine. It should be noted that the date of occurrence of the landslide is not explicitly received.

The determination of a predictive model of slope instability in the SE sector of the mine was made, based on the behavior of the records coming from the monitoring of prisms between 2014 and 2016, the variations of the DTM and their intersection with the block model. We remark that it was possible to determine predictions 2, 4 and 6 months ahead of an instability that should have occurred in the month of April 2017.
 
The selection of the forecast months is a definition; product that the information collected is defined as moderate, by the amount of prisms existing in the sector. With greater information in quantity and quality, the predicted period could be longer than the 6 months exposed in the analysis, in which new variables from other instruments could be included. To construct our predictor, several algorithms 
were tested: Suport Vector Machines, Neural Networks, 
 Bayesian Networks,  ICA,  CART and  Random Forest. All with different results and performances. The best one was Ramdom Forest who was agood predictor gives us a good discrimination and classification of the variables.

\begin{figure}
\begin{center}
\includegraphics[width=5in]{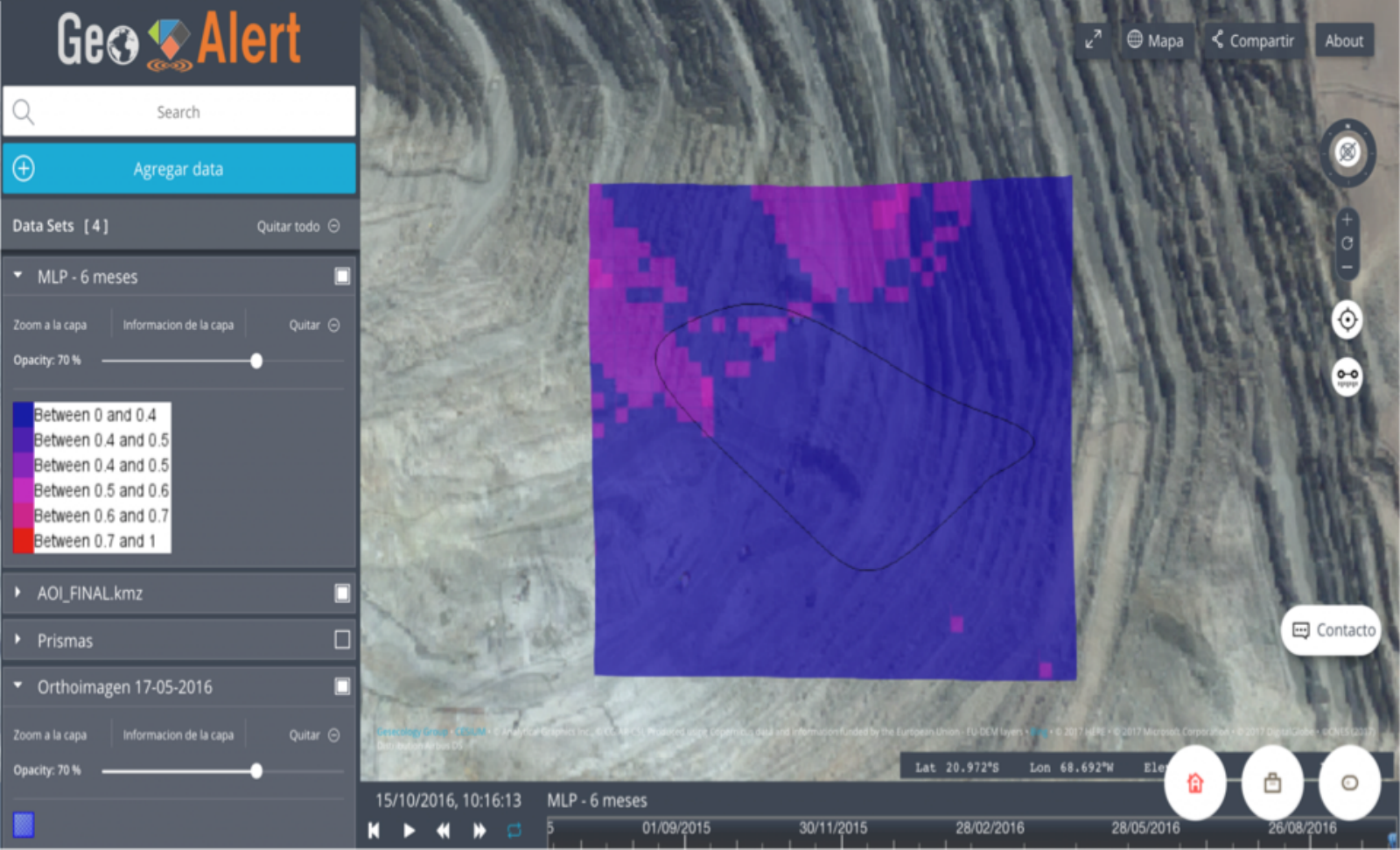}
\caption{Results of the risk analysis model in the study area}
\label{Fig6}
\end{center}
\end{figure}

Figure \ref{Fig6} shows the result of the long-term model calibrated for 6 months in advance, where a forecast for the month of April 2017 is shown, which shows an instability under our study region, an event that occurred according to information delivered later.

\begin{figure}
\begin{center}
\includegraphics[width=5in]{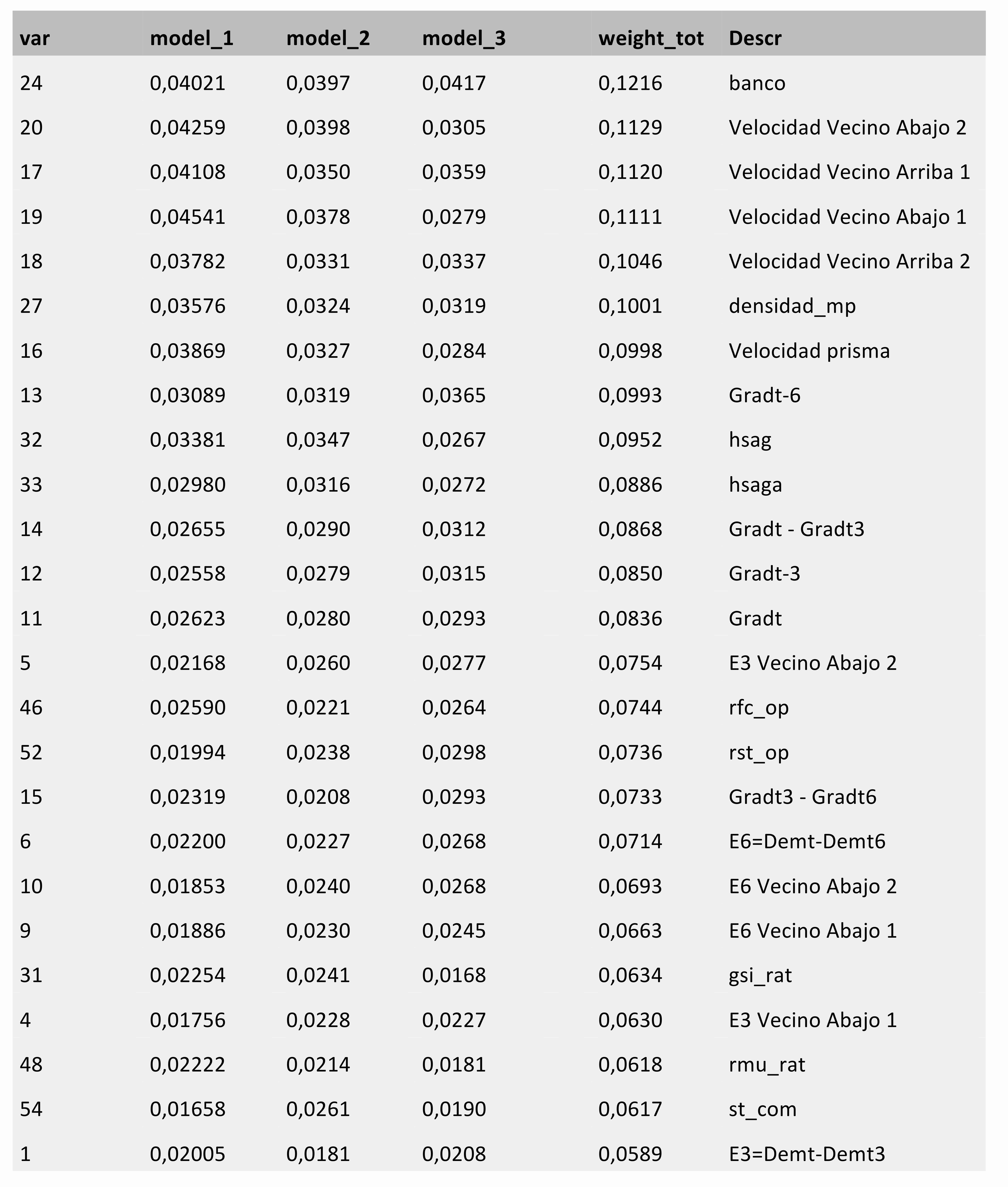}
\caption{Clasification of the used variables}
\label{Table1}
\end{center}
\end{figure}

It should be noted that a product that generates the execution of our risk analysis model is the classification of the variables, according to the influence on the forecast. This classification depends solely on the technique of data mining and the information provided by the mining company, it is important to note that this classification may vary depending on the data used, since the competition of the land or other variables vary greatly depending on whether the areas to study, such as considering different mines and geographical and climatic factors. As detailed above, different data mining techniques were applied for the analysis of information, the most effective in this case being the so-called {\it Random Forest} or also known as {\it decision trees}. Figure \ref{Table1} shows the results obtained according to the classification obtained, this classification is a product of data mining using Random Forest algorithm, which is not only a prediction tool, but also gives a classification of the variables under study.

\begin{figure}
\begin{center}
\includegraphics[width=5in]{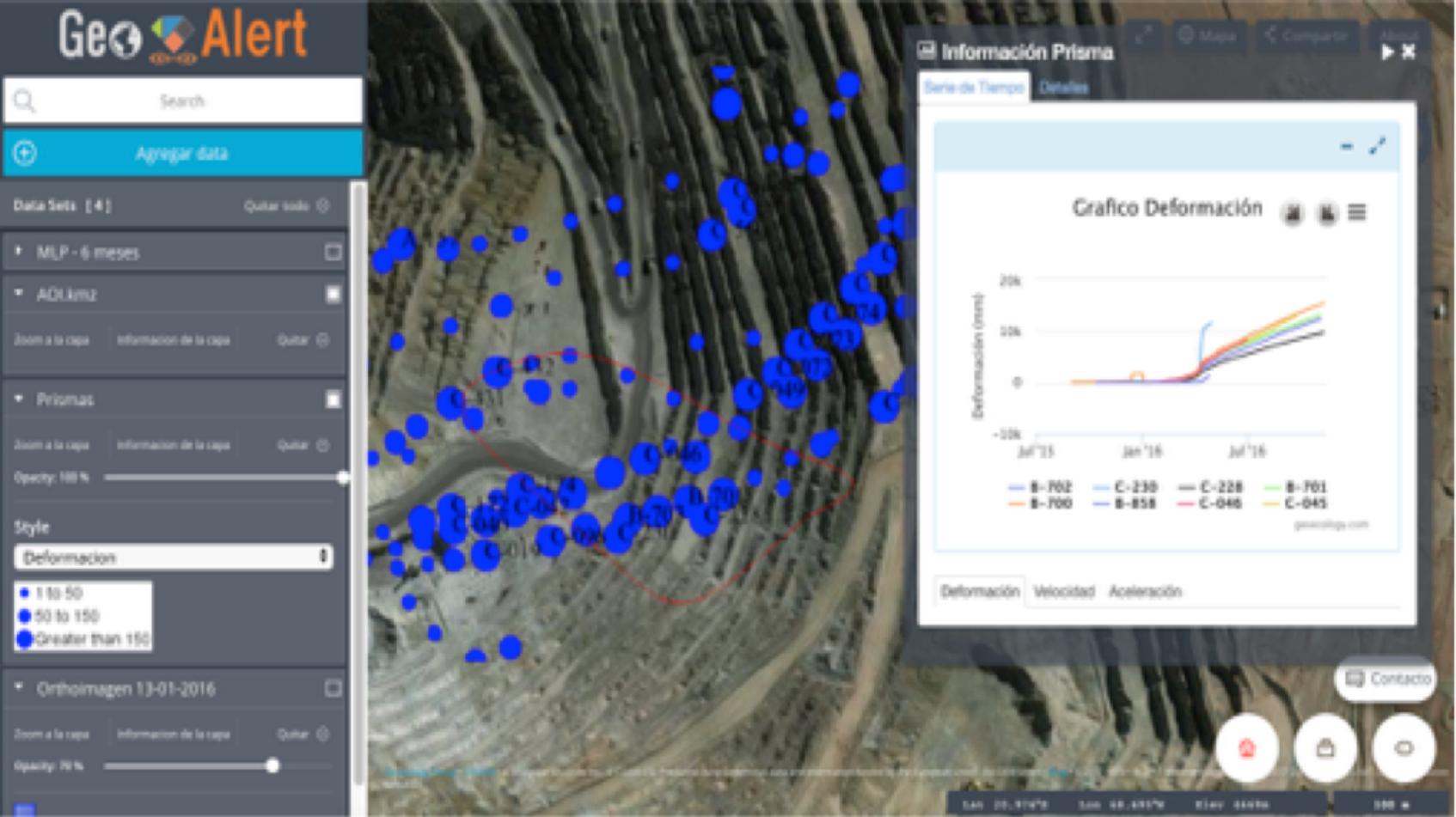}
\caption{Prisms information showed on GEOALERT system}
\label{Fig7}
\end{center}
\end{figure}

\ref{Fig7} shows the graph of the magnitudes of the deformations of the prisms registers in the Geoalert system in the SE sector, indicating that presumably the slip in the sector occurs during the month of April 2017. It should be noted that the date which is ratified with the prediction of 6 months resulting from the model. We can also observe that the landslide occurs under the study area and we can only observe part of it.

We can conclude in this way that the predictive model delivers a satisfactory result, according to the information processed. It stands out, the importance of having found reasonable predictive correlations considering only the historical behavior of only one type of sensor, as are the prisms. It should be noted that to achieve this result, a calibration and training of the model with part of the information delivered was needed. It stands out, the participation in the process, the basic information additional to the prisms, which corresponds to the variation of the DEM in the sector and its intersections with the block model, processing that is of three-dimensional character achieved thanks to the Geoalert platform. We can also note that the use of all the variables inserted in the block model, which means an important volume of information, from which it can be observed that the information obtained there is relevant for the study and which is key for the development of the predictive model. 
The received satellite monitoring reports could not be used because the information was only at the PDF file level, however, these reports do not show instability in the SE sector, which is opposed to the prism information, however it would be interesting in later stages the inclusion of this variable can already be an important element in the development of the predictive model. For this, and as a recommendation, to achieve superior prediction results in some other sector, additional monitoring systems should be considered, such as terrestrial and satellite radars and pore pressure information.

\section{Conclusions}
\
\par

As a first conclusion is that the risk analysis model used managed to detect the risk zones and in which the landslide occurred, which shows the efficiency of our methodology. The forecast horizon is related to the historical data and the event that will be used to perform the calibration of the Models.

It is important to note that the system will have a better prediction as long as there is a greater and better quantity and quality of the data used for the training of the system, which allows a better learning and the calibrations will be of better quality. However, the system is able to predict and/or generate risk indicators with a limited amount of data.

It is also important to note that as the volumes of data grow, it is possible that an adequate infrastructure is required for its management and processing, which includes high performance computing (HPC), but this is transparent for the user who makes use of the platform for viewing information. We also noted that in the development of this study, different data mining techniques were used, which were tested for different purposes, such as prediction or classification of the variables. Finally the method {\it Random Forest} showed a better performance in the achievement of a good prediction.

 As it was stated in this document, geographic and geological information is of great importance in the development of the model, however, its high cost allows a limited amount of drilling, so the determination of where to conduct the drilling is a critical decision in operations. In this sense, this tool, when determining the areas with the highest risk of landslides, can help determine the sectors where to conduct the drilling, prioritizing the areas of greatest risk and thus helping the planning and also making optimal use of available resources. We can also note that while 6 months of delivery time was used for the delivery of results, this time interval may vary, depending on the available information, however it should be noted that the efficiency would also depend on the available historical data.

In the same way, the choice of the location of the sensors is also a critical problem, and that has also been widely studied from various scientific aspects, so the determination of the risk areas, can help optimize the installation of The network of sensors, such as ground radars, high-cost equipment, and having more information would help in planning the installation of the sensor network.
The possibility of using the already calibrated Models and Forecasting System is now open, as a Simulation method using as Data, achieving as a result a Map of risk zones.

\end{document}